# On the role of Boundary Condition on the Speed- & Impact- Distributions in Dissipative Granular Gases in Knudsen Regime Excited by Vibration

**P. Evesque**
Lab MSSMat, UMR 8579 CNRS, Ecole Centrale Paris
92295 CHATENAY-MALABRY, France, e-mail: evesque@mssmat.ecp.fr

**Abstract:**
*Recent experimental results on granular gas in Knudsen regime excited by a vibrating piston in micro-gravity have measured distribution p(I) of impacts I with a fix target. They give p(I) ∝ exp(-I/I$_o$). This distribution leads to a probability distribution function of speed v along z varying approximately as f(v) ∝ (1/v) exp(-v/v$_o$); hence it diverges as 1/v at small speed and it is quite non Boltzmannian at large speed. Here, a model is proposed, which explains these experimental impact distributions; it takes account of the true role of the boundaries and of the dissipation in the gas. This validates the experimental data. Different approximations are discussed. Roles of boundaries and of 0-g condition are investigated theoretically. It is argued that the piston plays the role of an impact generator or a "velostat" for the Knudsen gas. These results cast a doubt on the efficiency of the notion of Boltzmann temperature and on the necessity to refer to Boltzman distribution in such dilute systems. The model shows also that the medium has to be considered as a whole, in global equilibrium: each part of the system is exchanging with the whole (at least in the direction of vibration); this is quite different from classic approach of dissipative systems based on local exchange and equilibrium, which leads to a "diffusive" Boltzmann equation; here the distribution f(z,t) is mainly propagative, i.e. f(z,v,t+δt)=f(z-vδt,v,t) instead of diffusive.*
**Pacs # : 05.45.-a, 45.50.-j, 45.70.-n, 81.70.Bt, 81.70.Ha, 83.10.Pp**

Many papers are dealing recently with dissipative granular gases, because these systems are expected to exhibit rather unusual behaviours. On the other hand, many of them start settling the problem using concepts from classic non dissipative systems, such as temperature, …., which may not be adapted. Most of them also neglect analysing or determining the role plaid by boundary conditions. However, we know from recent experiments how non extensive the physics of granular dissipative gas is. In this note, we just want to exemplify how much important these two parameters are.

We study a granular gas in a Knudsen regime, and we show how far its speed distribution is from Boltzmannian. We interpret it as linked to a coupling between (i) a specific boundary effect and (ii) specific propagation processes. For this purpose, we use data of [1, 2] that concern the experimental impact distribution in granular gas excited by vibrations; these data are reanalysed and interpreted within a simple model. In § 5.3 & 5.4 of [1], some suspicion was cast on the validity of these data, because they were demonstrating the existence of a **diverging probability of getting particles with very slow speeds**, and because no simple explanation was found. Here, a model is proposed to understand the results, which starts from the one proposed in §-5.3 of [1]; it is modified to take into account the real time a ball stay with speed v.

Essentially, the model is based on two processes, which are (i) the propagation of the ball at speed v and a speed increase during ball-piston collisions and (ii) efficient losses due to ball-ball collisions. This gives the trends observed experimentally. Hence this interpretation, which is simple, strengthens the validity of the experimental data, and confirms the correct working of the gauge. A description using Boltzmann equation is proposed. A discussion about the role plaid by the boundary is then undertaken. The notion of random impact generator is then introduced. This allows undertaking a discussion upon the differences between experiments on granular gases performed in 0-g and 1-g conditions.

These results cast a doubt on the efficiency of the notion of Boltzmann temperature and of Boltzman distribution in such dilute systems. They show also that the medium has to be considered as a whole system in global equilibrium in which each part of the system is exchanging directly with all the other parts, at least in the direction of vibration; this forbids the use of classic approach based on local exchange and equilibrium. (Such an approach is used commonly for other dissipative systems, such as viscous fluids or electric resistance; it is based on local description of the evolution of small volumes in interaction with their neighbours using Boltzmann equation of evolution; this one describes the exchange of matter, of impulse and of energy between local adjacent small volumes in partial equilibrium). The proposed model imposes (or takes into account) then the non extensivity of the physics and uses speed distributions which are propagative $f(r,v,t+\delta t)=f(z+v\delta t,v,t)$.

The problem can be settled as follow: the results which are being discussed are those ones of Fig. 9 of ref [1] and are reproduced here in Fig. 1. They have been obtained during micro-gravity conditions in Airbus A300-0g of CNES with balls (diameter d=2mm) in a fix cylindrical container (diameter D=13mm, Length L=10mm) closed on bottom by a vibrating piston, moving as b cos($\omega$t), and closed on top by a gauge. The gauge is at rest in the lab frame; it counts and measures the ball impacts I as a function of time. The number N of balls in the cell is small (N=12, 24, 36, 48) so that the gas is in the Knudsen regime, with a number $n_l$ of layer covering the bottom at rest $n_l=Nd^2/D^2$ ($n_l$=0.3, 0.6, 0.9, 1.14). One observes an exponential distribution for the impact amplitude (*cf.* Fig. 1). As a matter of fact some doubt was cast on the correctness of these results in [1], where it was observed that they indicate a probability distribution of speed v along z which diverges when v approaches 0.

The basic reasoning was as follows (*cf.* discussion on p. 27 of [1]): be p(I) the probability distribution function (pdf) of getting an impact of magnitude I=mv(1+ε), with ε being the restitution coefficient, and be f(v) the pdf of getting a ball with speed v, one shall get:

p(I)dI=vf(v)dv with dI=m(1+ε)dv (1)

As p(I) is found to decay exponentially according to A exp(-I/$I_o$), this predicts that f(v) varies as (B/v) exp(-v/$v_o$), which diverges as 1/v when v tends to 0, which means in turn that most of the particles are nearly at rest. This 1/v divergence indicates a behaviour which is far from the behaviour of classic gas. The exponential tail is also far from the classic 1d Boltzmann law $\exp\{-mv^2/(kT)\}$.

However we want to show here that this result is quite compatible with the physics, even if it looks quite anomalous when tenting to apply concepts of physics of classic gas theory to dissipative gases. The paper is organised as follows: section 1 derives a 1d modelling. Confrontations with experimental data are performed in section 2, while section 3 is devoted to analyse the effect of a non linear relation between the impact I and the ball speed. Then a 3d modelling is proposed in section 4 and a discussion on the role of boundary conditions is performed in section 5, with special emphasis on the role plaid by gravity and on the sensitivity of the experiment to a change of shape of the excitation law.

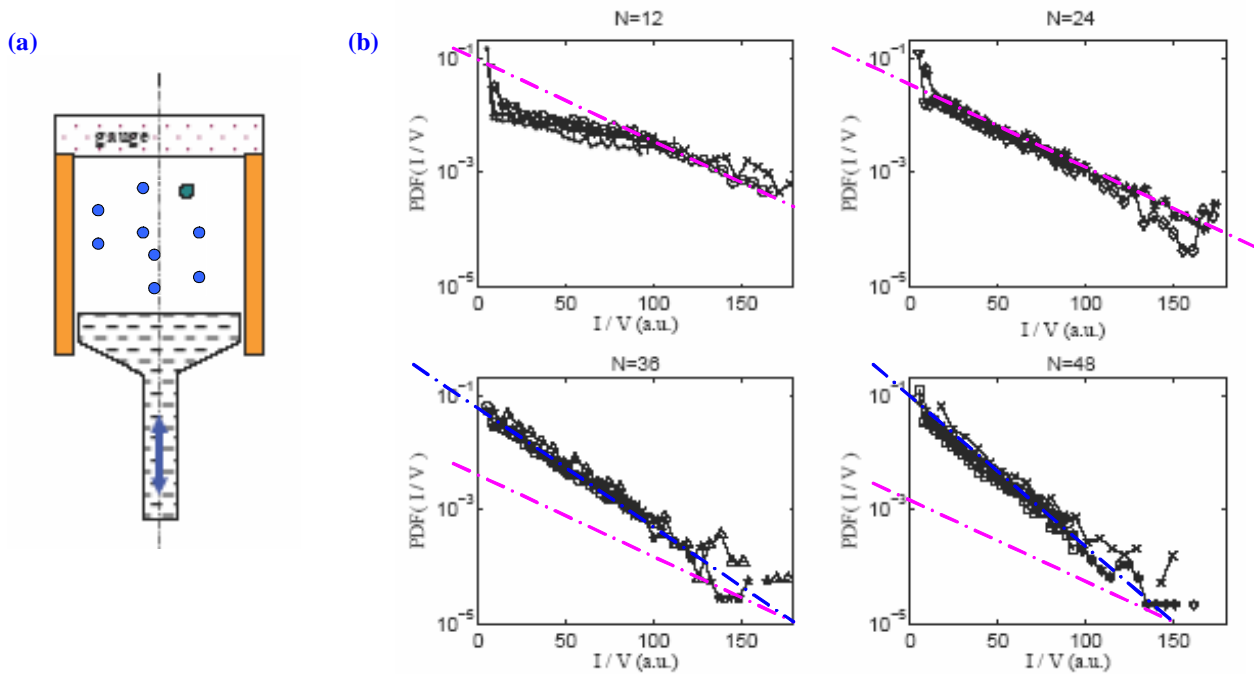

*Figure 1: (a, left): the experimental set-up is flown in Airbus A300-0g of CNES. (b, right) Probability density functions of the impact amplitude I measured by the sensor, for different vibration parameters during 16 s of low gravity, for different number of balls: N=12, N=24, N=36, N=48. Symbols are N=12 [#1 (x); 2 (○); 3 (●); 4 (+)]; N=24 [#5 (*); 6 (◊); 7 ( )]; N=36 [#8 ( ); 9 (pentagrams); 10 (○); 11 (x)] ; N = 48 [#12 (x); 13 (hexagrams); 14 (○); 15 (□)]. One can rescale all the curves into a single one, using the parameter $I\,N^{0.8\pm0.2}/(b\omega)$, cf. [1,2,3].*

## 1. Derivation of the exponential law of impact probability from a 1d model :

We start with the model proposed on p. 28 of [1]; we assume in this section that the impact I is proportional to the impulse mv, *i.e.* $I \propto mv$. The model states that the only way to get a ball with large speed $v \approx k\,b\omega$ is to find a ball which has hit k times the piston successively without hitting any other ball; so this probability scales as $\{\exp[-2(L-d)/l_c]\}^k = \exp[-2k(L-d)/l_c] = \exp[-8kn_l)]$, where $l_c$ being the mean free path between two ball-ball collisions, *i.e.* $l_c = D^2(L-d)/(4d^2N) = (L-d)/(4n_l)$.

However, what was missing in the model is the following: since the travel lasts $2(L-d)/v$ before a change of speed, the typical time spent on this configuration at speed v scales as 1/v. So Eq.(8) of ref. [1] shall be divided by v to get the right probability, because the configurations are time weighted; this leads to

$$f(v) = (1/v) \ \exp(-v/v_o) \quad (2)$$

As the number of impacts at speed v on a fix gauge varies as v f(v) and since the momentum transfer I is I=mv(1+ε), the distribution g(I) of momentum corresponding to impacts with the gauge shall vary as:

$$p(I)=\exp(-I/I_o) \quad (3)$$

This predicts the exponential behaviour of Fig. 1. So this simple model is compatible with the experimental data contrarily to what was thought in paper [1]. It may catch the essential physics, and it predicts the $I_o$ value to scale as

$$I_o \propto m v_o (1+\varepsilon) \quad (4.a)$$

with $\quad v_o = b\omega/(8n_l) \quad (4.b)$

and $\quad n_l = N\pi d^2/S = Nd^2/D^2 \quad (4.c)$

As a matter of fact, the factor 8 in Eq. (4.b) is an approximation; it is an average of all possible impact values that depends on the phase of the ball during impact, and on the speed [4]; this point will be discussed in sections 3 & 4.

## 2. Discussion:

At this stage it is worth discussing few experimental facts and theoretical approximations:

♦ One can see in the cases N=12 and N=24 of Fig. 1 an excess of collisions at impacts near v=0, compared to the exponential law. This excess corresponds to about 10-20% of the total number of impacts in these cases. This excess may be explained by the fact that particles with slow speed v, *i.e.* v ≪ bω, cannot gain so much speed when hitting the piston, because they hit the piston at its maximum elongation only, *i.e.* when its speed is quite slow (see [4] and Fig. 10 of [1]). Hence the speed gain of such sates is much smaller than bω. This reduces the probability of escaping from these slow states, increasing the duration of such states in turn, and the probability of finding them. The main way to escape from such a slow state in the case of the density range studied in [1] is that the ball collides with a faster ball, which arrives every $\tau_c$, where $\tau_c \approx l_c/\langle v \rangle$ is the typical collision time, and ⟨v⟩ the mean ball speed.

♦ One sees also in Fig. 1, N=12, that a plateau exists before the exponential decrease. This occurs in the slow speed region; we have no explanation for this trend at the moment.

♦ On the other hand, noise on signal from the gauge exists; it leads to added small peaks; it means that the smaller peaks can be due to noise and are artefact. To which extent this perturbs the results of Fig.1, one does not know at the moment; but it introduces likely a cut-off $I_{min}$ at small impact, that may depend on b, ω and N.

However large peaks are not generated by noise; so the exponential decay at large impact is experimental indeed, while the peak of density near I=0 is perhaps not.

♦ As a matter of fact the collision of a slow ball with a faster ball shall results in the increase of the speed of the slow ball. This process is neglected in the above model since ball-ball collisions are supposed to slow down efficiently the speed only. The approximation, which neglects speed increase by ball-ball collision, may be good enough however, even in the range of intermediate speeds of Fig.1, because the speed gain due to a ball-ball collision is distributed at random along the three directions (x,y,z), and because the only one which intervenes in the model is the contribution in the direction of propagation Oz ; hence, this limits efficiently the discrepancy; (the x and y transfers do not contribute to I).

♦ Nevertheless, the above model should apply preferentially to speedy balls, which have performed few roundtrips without hitting a ball. So it should apply for speed v large enough, *i.e.* v > 3 bω or so. The typical speed <v> observed in these experiments is bω about according to Table 1 data of [1] and or in Fig. 7 of [1] and reported here in Fig. 2; hence it corresponds only to one roundtrip!

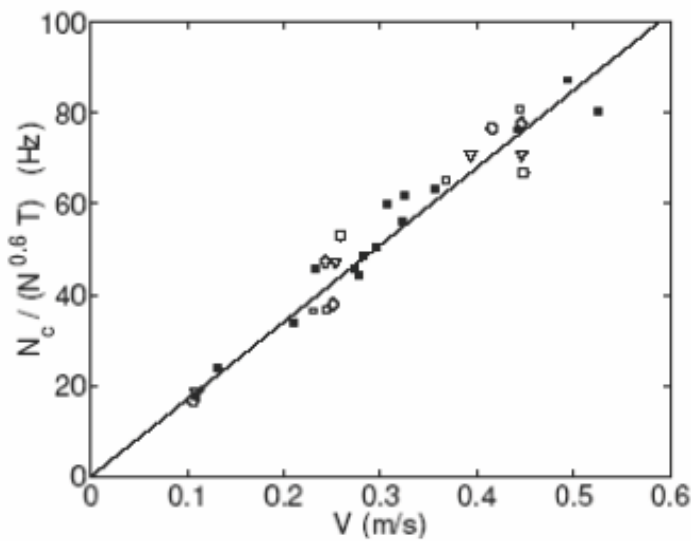

*Figure 2:*

*Total number of collisions $N_c$ observed during time T, rescaled by $N^{0.6}$ and by time, $N_c/(T\ N^{0.6})$, as a function of V=bω for N =12 balls (□) and (■); 24 balls (♦); 36 balls (∇); 48 balls (o) during T = 16 s of low gravity. ■ marks are from experiments with N=12 balls and 15 different velocities (cf. [1]). From [2].*

*N is the number of particles in the cell. Solid line corresponds to the fit $N_c/(T\ N^{0.6}) = \alpha V$.*

Hence there is some discrepancy and inadequacy between the model and the experimental data. Nevertheless, they look in rather good agreement; so we may think the model describes approximately the features and that it explains also the divergence of the number of balls at small speed. Hence this comforts the experimental results and raises away the doubt on their validity. So, one can handle the data for themselves and try to extract what they exactly tell. This will be done in the next section; before, it is worth noting the following:

♦ ***Effect of a velocity-dependent gain:*** Be v the speed of the ball which generates an impact of intensity I, be v+Δv the speed for an impact intensity I+ΔI and suppose v and I to be proportional. As a matter of fact, one can remark that Eq. (3) tells simply that the occurrence probability of the impact I+ΔI, is just the one of getting the impact

I multiplied by $P_o=\exp-(\Delta I/I_o)=\exp(-\Delta v/v_o)$. Indeed as speedy enough particles can only gain velocity by impacting the piston without hitting an other ball; hence $P_o$ is just the probability to travel a length $L_{\Delta v}$ necessary to accelerate from v to v+$\Delta$v; this can occur if the ball hits the piston a sufficient number of times without hitting any other ball meanwhile.

Suppose now that the speed gain $\Delta_1 v=a_1 b\omega$ per piston hit is independent of the ball speed v, this imposes that $P_o$ scales as $P_o=\exp(-L_1/l_c)$, with $L_1$ proportional to $a/a_1$ and to L, so that

$$P_o=\exp\{-\Delta vL/(a_1 b\omega l_c)\} = \exp(-\Delta I/I_o) \tag{3}$$

This is a new way to derive the result, which does not need using the distribution f(v). Both ways are equivalent. This last one allows discussing the effect of a gain of speed which would be speed dependent.

Indeed, one expects also that the speed gain G(v) per piston hit depends on v [4], because on some phase effect; hence one shall expect also some deviation from the exponential law:

- For instance for $v \leq b\omega$ about, one expects that the mean gain increases with v so that the log(I)-*vs.*-I curve shall be concave (positive curvature, *i.e.* $\partial^2[\text{Log}I]/\partial v^2 > 0$, *i.e.* the slope of Log(I) *vs.* I decreases at large I, since $\partial[\text{Log}I]/\partial v < 0$).
- On the contrary, for $v \gg b\omega$, $v<v_{max}$, [5] the gain decreases with v tending towards 1 at large speed, so that the log(I)-*vs.*-I curve shall be convex in this range (negative curvature, *i.e.* $\partial^2[\text{Log}I]/\partial v^2 < 0$).

In fact, the gain could be negative too, due to dissipation during piston-ball collision for very large v, *i.e.* $v>v_{max}$; but this range shall never be observed experimentally, because it is quite improbable under usual condition.

An other point which is worth noting is the fact that impact amplitude I may vary non linearly with v. This occurs for instance in a Hertzian contact when I is proportional to the maximum applied force; in this case, I scales as $v^{4/5}$. It results in an increase of the apparent non linear gain of $\Delta$I *vs.* I; this increases also the positive curvature (in the case of Hertzian contact due to the 4/5 power law relation). This is studied now, in the next section.

## 3. Including I vs. v non linear behaviour:

Calibration of the sensor has been performed [6] using a 1-ball experiment as in [1]. Various piston shapes, ball sizes and different materials have been tested. It was found that the impact amplitude I and duration $\tau_c$ varies according to a law of Hertzian contact:

$$I \propto v^{4/5} \tag{5.a}$$

and

$$\tau_c \propto v^{1/5} \tag{5.b}$$

Fig. 1 experiment tells that p(I) varies as exp-(I/$I_o$) approximately, and that $I_o$ depends on bω and N as bω$N^{0.8}$. This trend can be expressed in terms of the speed v, just by changing of variables from I to v using Eq. (5.a), *i.e.* I/$I_o$=(v/$v_o$)$^{4/5}$ . So calling ρ(v) this distribution of speed v responsible of the distribution of impacts, one has p(I)dI= ρ(v)dv, which leads to p(I)= ρ(v)/{∂I/∂v} and to:

$$\rho(v) = \rho_o \, v^{1/5} \exp\{-(v/v_o)^{4/5}\} \tag{6}$$

with $\rho_o$ being a normalisation factor. From a theoretical view point, one expects the previous model to be approximately valid, which predicts that $v_o$ varies according to Eq. (4.b) . In turn this predicts that $I_o$ shall vary as

$$I_o \propto (b\omega)^{4/5} \, n_I^{4/5} \tag{7}$$

Experimentally, it is found $I_o \propto b\omega N^{-0.8}$. So the predicted scaling for $I_o$ *vs.* N agrees with experimental ones, but the $I_o$ *vs.* bω predicted scaling differs slightly for the experimental one.

As the calibration shows that v varies faster than I, *i.e.* I ∝ $v^{4/5}$, the present theory which predicts an exponential decay for the distribution of impulse mv, predicts also a convex behaviour, with small convexity in the ln(p)-*vs.*-I representation, since I∝ $v^{4/5}$. This is not observed on Fig. 1. So there is a slight misfit between the theory and the observation, due to the non linear behaviour existing between I and mv.

## 4. Starting from 3d modelling to get the above 1d description:

All what has been told previously has been analysed using a 1d modelling. But the problem is 3d. So, is it valid? The aim of this section is to derive the 1d model from the 3d problem. Owing to the cylindrical symmetry of the geometry, it is better working with axial coordinate. We label v the speed coordinate along z, and $u_x$,$u_y$ (or $u_r$, $u_θ$) the transverse velocities. Be f(v,u,r,z,t) the probability distribution of speed at position (r,z) and time t for one particle. We expect, *i.e.* hope, that the particle motion is erratic enough to loose correlation with one another rapidly and that the system is periodic with period T. Then averaging over a period and a section of the cell, one may hope the system stationary, ∂f/∂t=0, with the new distribution f :

$$\underline{f}(v,u,z) = (1/T)(1/S) \int dt \int dx \, dy \, \underline{f}(u,v,x,y,z,t). \tag{8}$$

The experiment consists in measuring p(I) at the gauge position z=0. For sake of simplicity, let us assume that I is proportional to v, *i.e.* I=$m_I$v where $m_I$ has a dimension of mass ($m_I$ is not the ball mass, but $m_I \approx (1+ε)m$). Then one gets:

$$p(I) \, \delta I = (1/T) \int \underline{dt} \int_v^{v+\delta I/m_I} v dv \int \underline{f}(v, u_r, u_θ, t, z=0) \, u_r^{d-2} du_r \, du_θ \tag{9}$$

So, labelling f(v,z) the distribution after integration on u and t, one gets:

$$p(I) \, \delta I = \int_v^{v+\delta I/m_I} v \, dv \, \underline{f}(v, z=0) = (I/m_I^2) \, f(v=I/m_I, z) \, \delta I \tag{10}$$

Choosing small enough δI leads to f(v,z) that does not vary much in between the integration limits v and v+δI/$m_I$. This leads to:

$$p(I) = (I/m_I^2) \; f(v=I/m_I, z) \tag{11}$$

Similar (but different) relation can be found if the law relating v to I is not linear.

## *Mean gain and Dispersion of gain:*

Within this model, particles are assumed to circulate along z all the time. Hence, the evolution of p(I,z=0) versus I can be calculated by a perturbation theory in the limit of large I, *i.e.* when $\int_v^\infty f(v,z)dv$ is small enough to be negligible, so that the probability of a collision of a ball at speed v with a faster ball is small and can be neglected. In this case, large impacts between a ball and the fix gauge result essentially from former collisions of the same ball with the gauge at a slightly different speed.

So, considering a ball of speed v, which impacts the gauge at time t, this ball has impacted the piston some time t' earlier with a speed v', and has impacted the fix gauge at some other time t", earlier than t', with speed v". Owing to propagation rules and collision rules, one gets :

$$t'=t-[L-d-b\omega \cos(\omega t')]/v \tag{12.a}$$

$$t''=t'+[L-d-b\omega \cos(\omega t')]/v' \tag{12.b}$$

$$(-v'\varepsilon_p) = v+(1+\varepsilon_p)b\omega\sin(\omega t') = v''\varepsilon\varepsilon_p \tag{12.c}$$

where ε and $\varepsilon_p$ are the restitution coefficient with the fix gauge and the piston respectively.

So, v' and v" depend on t' and v. Also, the probability the ball hits the piston depends on t' since it depends on the relative speed v'-bω sin(ωt'). So the value of v" (and v') which is able to generate v after a roundtrip is not unique, but is spread over some range which depends on v, with a distribution of probability.

A way to estimate the distribution of v' and of v" which corresponds to v is to use the random phase approximation (RPA) [4], for which the balls are assumed to arrive at any fixed position z at random time, *i.e.* without any correlation between the balls themselves and/or with the phase of the piston motion. A way to verify the validity of this approximation is to measure the distribution of waiting times between two consecutive collisions of balls with an immobile target; it shall exhibit an exponential distribution if it occurs at random. This is just what is observed with the experiments performed in the Airbus A300-0g of CNES [1] which are reproduced here after in Fig. 3. So we know that this approximation looks satisfied in the present case [1].

However, when considering collisions with a mobile target such as the piston, and because the piston speed varies with time, one has still to take into account the fact that the probability of hitting the piston varies with the relative speed; hence this probability varies as

$$Pdt'= [v'-b\omega \sin(\omega t')]dt' / \int dt'[v'-b\omega \sin(\omega t')] \tag{13}$$

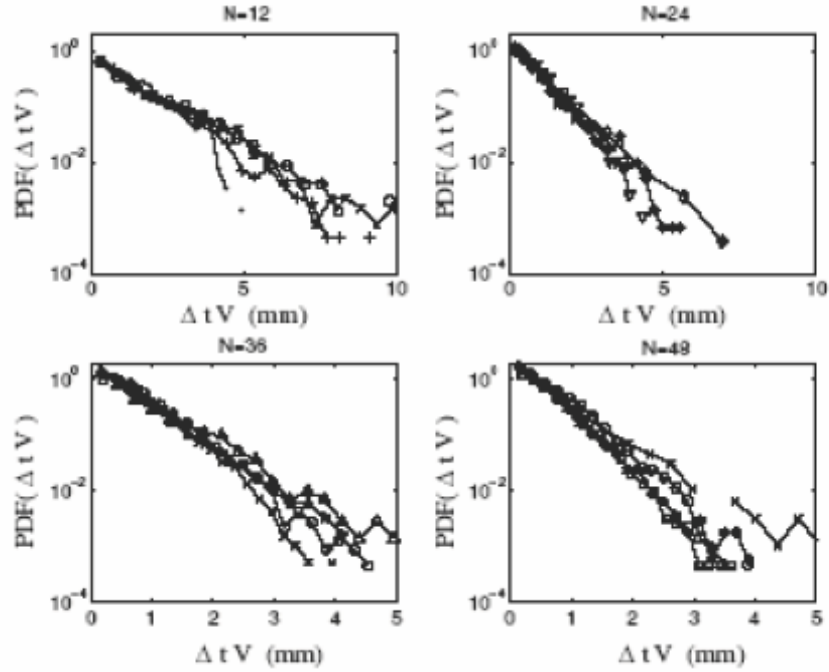

**Figure 3 :** Probability density functions of the free time Δt separating two successive collisions with the fix gauge, rescaled by V (V =bω), for different vibration parameters during 16 s of low gravity: N = 12 [#1 (x); 2 (o); 3 (●); 4 (+)] ; N = 24 [#5 (*); 6 (◊); 7 (∇)] ; N = 36 [#8 ( ); 9 (pentagrams); 10 (o); 11 (x)] ; N = 48 [#12 (x); 13 (hexagrams); 14 (o); 15 (□)]. From [2], see table 1

In Eq. (13), the integral has to span over a period T, and it is assumed that v'>bω (more complicated integral limits have to be used in the case v'<bω as shown in [1], so that we restrict the investigation to v'>bω here after for sake of simplicity). This leads to:

$$\text{Pdt'}= [v'-b\omega \sin(\omega t')]dt'/(v'T) = dt'/T \{1- [b\omega/v'] \sin(\omega t') \} \qquad (14)$$

From Eq. (12.c), one gets $(b\omega/v')\sin(\omega t')=-(\varepsilon_p+v/v')/(1+\varepsilon_p)$ and $dt'=-dv'\{\varepsilon_p/(1+\varepsilon_p)\}/\{b\omega^2\cos(\omega t')\}$. According to this, the ball which hits the gauge with speed v was then issued from an earlier collision with the same gauge at time t"=t-(L-d-bω cos(ωt'))(1/v-1/v')= t-(L-d-bω cos(ωt'))[1/v+1/(εv")], with a speed v"=-εv', where $\varepsilon_p$ and ε are the normal restitution coefficients of the piston and of the gauge respectively. This occurs if the ball does not hit any other ball meanwhile, which probability is exp[-2(L-d- bω cosωt')/$l_c$]. So, approximating exp[-2(L-d-bω cosωt')/$l_c$] by exp[-2(L-d)/$l_c$] one gets:

$$vf(v)dv=p(I)dI= \exp[-2(L-d)/l_c] \int g(v",v) \, v"f(v")dv" \qquad (15)$$

where g(v",v) is the probability that ball with the speed v" is transformed into a ball with speed v after a roundtrip, which includes a collision with the gauge and with the piston, and which is conditioned that the ball has not hit an other ball meanwhile. So, g(v",v) is given by Eq. (14) in which v, v' and v" are related by Eq. (12):
     This fixes the range of v which can be generated from v"; it fixes also the time t' at which v' collision between the piston and a ball at speed v' shall occur to get

v"→v'→v ; it fixes at last the laps of time dt' to be considered for a given dv' and dv", since dv" is related to dv', and since dt' is related to dv' at constant v. This makes the problem of determining g(v",v) solved. As we have assumed also in the present section that v=I/$m_I$ whatever I and v, this leads to:

$$p(I)dI = \exp[-2(L-d)/l_c] \int g(I''/m_I, I/m_I) \, p(I'')dI'' \qquad (16)$$

Eq. (16) tells that gain v-v" is not constant for a given v, but depends on the time (or phase) at which collision with the piston occurs. The function g describes the distribution of gain; its averaging gives the mean gain.

***Remark 1:*** It is worth noting that the probability $\exp[-2(L-d)/l_c]$ is an approximation of the probability that a ball does not hit an other ball; it is valid only when the considered ball is fast enough, *i.e.* when v > <v>, as noted already in point 1 of §-2, because when v < <v> , collisions with the considered slow ball occur with faster balls ($v_f$) most of the time since they travel much more in the same time. Better evaluation of this probability and its dependence on v/$v_o$ might be goal of further investigations.

***Remark 2:*** It is worth noting that the model is based on a Boltzmann's equation of evolution which is propagative and not diffusive since the *distribution f(z,v,t) is deduced directly from f(z',v,t=0) since f(z,v,t)=f(z'-vt,v,t=0)* while the dissipation is treated as a process which dissipate towards a bath whose feed-back contribution is neglected. Hence, we do not believe it is the usual way to write the Boltzmann's equation, which writes in general transfer due to collisions with adjacent volumes. In this sense the model is non local and uses **propagation equations** instead of diffusion equations and local couplings.

***Remark 3:*** It is worth noting also that the slow speed region of the distribution might be a goal for further investigations too, because the balls in this range are likely more coupled to their local environment, and less to boundaries. Hence their physics shall looks more similar to the one of classic gases; however they are still coupled to much faster balls which exhibit already strongly anomalous distribution. So this part of the distribution shall reflect how fast Boltzmann equilibrium distribution may be or may not be recovered due to the balance between (i) the local interactions which is expected to enforce a random Boltzmannian-like distribution and (ii) the forcing by the anomalous distribution.

## 5. Discussion on the role plaid by boundaries; effect of gravity:

From the experimental result, and from the modelling, it is clear that boundary plays the role of a generator of random impacts, when coupled to the granular gas, with a mean $I_o$ (and $v_o$) that depends on bω. It depends also on $n_l$=N/S since this parameter define the coupling with the gas. This fixes the internal dynamics in 0-gravity. This is indeed what is observed with the experimental results in Airbus A300-0g. And the impact distribution looks exactly as if the system was running as random as possible, keeping in mind that the mean impact $I_o$ shall be given, (for a given b,ω, $n_l$ =N/S) (one knows the most disordered solution correspond to the exponential law, (see [1] if necessary). Such a boundary condition can be called a "velostat" [7], since it tends to impose a typical velocity in the z direction, and since this typical velocity is related to the typical velocity of the wall bω. Also, many reasons exist that the true f(v) distribution may not be exactly exponential as discussed in §-2 & 3; it is imposed by the dynamics of the collisions, which is self adjusted.

● The typical time for a roundtrip $T_r$ is expected to scale as $\int L/v\, f(v)dv$, which may/shall diverge due to slow-bead contribution. However, it shall scale as $L/v_o$ about for faster beads. An other way to proceed to evaluate $T_r$ is to count the number $N_c$ of impacts in a given time T. One can calculate $N_c$ directly since $2LN_c = N \int Tv\, f(v)dv$, which converges. Then $N_c=NT/T_r$ that implies $T_r=T\, N/N_c$ and $v_o=<v>= 2L/T_r =2LN_c/(NT)$. Also, one can define the typical time τ between two impacts, which is $\tau=T/N_c=T_r/N$ :

$\quad \tau = 2L/(Nv_o)$ \hfill (17.a)

This time τ corresponds also to the mean time between impacts on a fix target; hence it can be measured from the distributions of Fig. 3 when knowing $v_o$. One gets

$\quad b\omega\, \tau = (2L/N)\, (b\omega/v_o)$ \quad or \quad $V\tau = (2L/N)(V/v_o)$ \hfill (17.b)

As it is expected from the model, the speed distribution f(v) shall be given when bω and $n_l$ are imposed. Then v does not depend on L. So the number of impacts per unit of time shall decrease as 1/L; it shall be also proportional to S at $n_l$ given, so to N at $n_l$ given. This is true as far as no clustering in lateral direction is formed.

Furthermore, as large speeds are generated by balls with large speed, correlations shall exist between large impacts within a typical laps of time $\tau_\chi = 2\, L/v_o$ about. However, as these correlations are internal to the dynamics of a single bead, they can be masked in experimental data by the dynamics of the other ones since all balls are moving independently from one another.

● The impact intensity I scales approximately as a percussion intensity Δ(mv). Hence $F\approx\int I\, p(I)\, dI$ is approximately the mean force applied by the gauge on the granular gas. It is a free quantity in weightlessness (0g), which results from the imposed dynamics, with a mean $I_o$. As the pressure P is given by $P= (1/S) \int \Delta(mv)\, p(\Delta(mv))\, d(\Delta(mv))$, it is expected to scale as $m(1+\varepsilon)v_o N/(ST_r)= m\, n_l\, (1+\varepsilon)v_o^2$, with $v_o$ which depends on $n_l$ at least.

● The amplitude b of vibration may play some role too, through the ratio b/L, because when b/L becomes larger than a given threshold α, the typical ball speed allows the ball to perform a round within less than a few periods so that resonance can occur. This will perturb the faster part of the speed distribution. $\alpha <1/\pi$ and depends on the dissipation, hence on ε, $\varepsilon_p$ and $n_l$.

### *1g experiment*

Do these results hold true in 1g? Obviously this is not certain. For instance, stationary conditions impose some different boundary conditions at bottom and on top; also, the top condition differs if there is a lid or not or if the lid is fix, or if it is a mobile mass $M_l$. Indeed, be $M_G$ the total mass of grains in the cell and g the gravity. So, in the case of a mobile lid, one shall have on top and bottom:

$$\int_{\text{on top}} I\, p(I)\, dI = (M_l)g \qquad \text{if the lid is mobile} \qquad (18.a)$$

$$\int_{\text{on bottom}} I\, p(I)\, dI = (M_l + M_G)g \qquad \text{if the lid is mobile} \qquad (18.b)$$

while, one shall get on top and bottom when the lid is fix:

$$\int_{\text{at bottom}} I\, p(I)\, dI = (M_G)g + \int_{\text{on top}} I\, p(I)\, dI \qquad \text{if the lid is immobile} \qquad (19)$$

And the $\int_{\text{on top}} I\, p(I)\, dI$ is free to adjusts itself to the physical conditions.

This defines the boundary conditions; they depend on g; so, the experimental data shall vary also with g. We defined also $v_o$ as the mean speed.

● When there is no lid, the typical roundtrip time shall scale as $T_r = v_o/\sqrt{g}$, and the faster the ball the longer its return time, just the opposite from the 0g! When there is a lid, the slower ball do not hit the lid, while the faster hit it; there is then a natural "cut-off" speed $v_c = (Lg)^{½}$ (where L is the height of the cell); $v_c$ is free compared to $v_o$, since it depends on L and g. When $v_c$ happens in the queue of the v-distribution, this shall perturb the end of the distribution, because the rapid balls are enforced to come back faster to the piston; their frequency are "artificially" increased. Also, as $v_c/v_o$ depends likely on bω, it may enforce some anomalous dependence with b and ω (or bω). Also, when b/L becomes large, resonance similar to the one observed in 0g can occur; and its threshold depends not only on b/L, but also on g and ω now. This makes the dynamics more intricate.

● When there is a mobile lid, the boundary condition is to enforce a constant transfer of momentum per unit of time; it is not obvious that this conducts to the same $v_c/v_o$ ratio, independent of b and ω:

Assume that $v_c/v_o$ remain constant, then L scales as $v_c^2 g$ and time as $T_c = L/v_c = v_c/g$. So if $v_c$ and $v_o$ scales as bω, the frequency of collisions with the lid as $1/T_c \approx 1/(b\omega)$. So the transfer of momentum $mv_o/T_c$ per unit time shall be independent of bω and the lid of constant mass shall remain in equilibrium at height L scaling as $b^2\omega^2/g$. On the contrary, if $v_c/v_o$ varies with the excitation, predictions are not clear.

● It is worth noting the role of slower balls which is quite different: these balls remain often in contact with the acting piston where they constitute a "dense" layer in 1g. In 0g, these slow beads do not often hit the piston and they are spread over the whole cell; this is important when local interaction starts becoming important. Also the impact distributions on the lid is different from the one on the piston, and they are both different from the one of any intermediate (fictive) surface.

### *Effect of the time dependence of the piston speed*

The function g(v",v) of the previous section, §-4, depends on the exact piston motion. For instance let us consider the case of a saw-teeth motion of the piston, with piston speed $V_+$ and $V_-$ such that $V_+T_++V_-T_-=0$ and $T_-+T_+=T$ and a ball at speed v"<$V_-$. The gain of such a system is unique for a given v", but varies with v": be v" the ball speed before the impact with the fix gauge, v' the speed after it, *i.e.* v'= -εv", (v' is also the ball speed just before the next impact with the piston) and v the speed after this second impact, one has: v'=-εv" and -$\varepsilon_p$(v'-$V_+$)= v-$V_+$, which leads to v=ε$\varepsilon_p$v"+(1+$\varepsilon_p$)$V_+$ , and to the gain ;²

$$v-v" = -(1-\varepsilon\varepsilon_p)v+(1+\varepsilon_p)V_+ \qquad (20)$$

Hence, the speed gain v-v" does not depend on the phase, so it is not distributed as it was with a sinus function; but it depends still on v, since it decreases slightly when v increases.

The gain reaches 0 for $v_{max}$=(1+$\varepsilon_p$)$V_+$/(1-ε$\varepsilon_p$) if $\|V_-\| > \varepsilon\|v_{max}\|$=ε(1+$\varepsilon_p$)$V_+$/(1-ε$\varepsilon_p$) . On the contrary, when $\|V_-\| > \varepsilon\|v_{max}\|$=ε(1+$\varepsilon_p$)$V_+$/(1-ε$\varepsilon_p$), the dynamics reaches a complicate regime because faster balls can hit the piston when it moves backward, and looses a lot of energy just in a single hit; The larger $V_-/V_+$ the scarcer these hits; this perturbs completely the dynamics and makes it chaotic likely; it turns to be some kind of Fermi's dynamics.

So there is a one-to-one correspondence between successive impacts of a ball subject to a saw teeth excitation, as far as no collision occurs with other balls. This differs from the sinus excitation. As the number of roundtrips increases the beads may loose energy through collisions with the cloud of particles. To which extend the global dynamics of the system is sensitive to a variation of the excitation profile, this remains to be determined.

## 6. Miscellanous: Corrections on references in [1]

The following references have to be amended:
[23] P. Evesque. "Statistical mechanics of granular media: An approach à la Boltzmann", *poudres & grains* **9** pp. 13-19 (15 November 1999). http://www.mssmat.ecp.fr/sols/Poudres&Grains/poudres-index.htm
In references [9 & 11] C. Chabot is now C. Lecoutre, since her wedding.

## 7. Conclusion:

This article proposes a modelling of the speed distribution in granular dissipative gas in a Knudsen regime and excited by a vibrating piston. It is based on a mechanical analysis of the problem, in terms of probability, energy gain and energy losses during collisions with piston and with the balls respectively. It takes into account the true role of boundaries and the role of dissipation of the gas. It predicts the exponential decay of p(I) *vs.* I, where p(I) is the probability of finding an impact of size I ∝ mv, when the gain of speed per piston speed is independent of v. Otherwise it predicts some deviation from this law.

The first step was to derive a mean field approximation; this has been done in two different ways, leading to the same result (§-1). Then a 3d modelling has been solved (§-4) which uses some kind of Boltzmann's equation formalism with long range coherence and using propagating function distribution; the method consists in looking for a stationary regime after averaging over a cell section and a period, and averaging also over a roundtrip of the ball; hence the creation term contain a combined process of (i) propagation and (ii) gain of speed which are (iii) combined to a conditional probability of non dissipation. So the approach takes directly account of multiple processes in a single shot; the particularity of the method is to take account of non local effect directly; it leads to an integral equation (Eq. 16) for large enough speed, which have to be solved self consistently and for each peculiar case.

The method can be improved by introducing new terms describing less important events such as those with ball-ball collisions. This would be a necessity when trying to describe the distribution of balls with slow speed.

The model agrees rather well with the experimental data, even if the range of parameters studied is slightly out of the zone for which the model applies, as shown in §-2. So it means that further investigation has to be conducted to improve the understanding.

Owing to the good correspondence between experiment and theory, it is worth to improve the comparison. So, due to the fact that some non linearity between impact amplitude I and ball speed v has been observed during gauge calibration, modification has been introduced to the model to take account for this non linearity (*cf.* §-3); it turns out that that the new corrected model agrees less with experiments than the previous one. So, this requires further investigation too.

Also some discrepancies between theoretical prediction and experimental data have been pointed out, which shall lead to further investigation. In particular, due to the fact that the gain of ball speed depends on the ball speed at which it hits the piston in the case of sinus excitation, one expects some curvature of the ln(p) *vs.* I law, *cf.* §-2; this is not observed, and requires more attention.

Finally, the effect of boundary can be and has been interpreted as playing the role of an impact generator or of a "velostat" when they are coupled to a granular gas in Knudsen regime.

As a conclusion, these results cast some doubt on the efficiency of the notion of Boltzmann temperature and of Boltzman distribution in such dilute systems. They show also that the medium has to be considered as a whole system in global equilibrium in which each part of the system is exchanging with all the parts, at least in the direction of vibration. This forbids the use of classic approach based on local exchange and equilibrium, which are used for instance for describing dissipation in viscous systems, or in electric resistance, whose description is based on the Boltzmann equation of evolution describing the exchange of matter, impulse and energy between local adjacent small volumes in partial equilibrium.

So, this comforts our previous intuition and understanding [7, 8]. Role of dissipation is important also for describing other effects, such as the Maxwell's demon [8]. In this peculiar case, however, the equilibrium in between the two sides of the slit is ensured by the transverse lateral motion, *i.e.* perpendicular to the vibration, whose dynamics requires ball-ball collisions. Hence, it is a slightly different case.

At last, when the excitation is not a vibrating piston but is caused by the cell which is moving periodically, the distributions of impacts in time and in intensity are both modified because they are measured by a gauge which is fix in the cell frame and mobile in the lab frame, *cf.* Eq. (13)... So one has first to correct the experimental data and/or the prediction from this artefact; this requires taking into account the motion of the gauge. Then comparison can be made efficiently.

Also, these data concern granular dissipative gas in the **intermediate** Knudsen regime. In the very low density regime, we have recently shown a completely different behaviour with the freezing of all rotation degrees of freedom and of the translation degrees of freedom perpendicular to the vibration [9, 10], leading to completely non ergodic behaviour; this effect is induced by the increase of dissipation due to solid friction when collision occurs with sliding contacts. This illustrates how much care has to be taken before tempting extrapolating the present model to denser or looser systems. This shows that dissipative granular gas exhibits a physics which is quite puzzling indeed. Are these new findings just the beginning of further multiple astonishments?

It is also surprising that no DEM simulation describes the above behaviours. This means first that these data and their interpretation have to be taken with caution and that they require more investigation to be considered as certain. They shall be also reproduced with computer simulations. But the fact that no simulation describes the observed results is not enough to deny their validity: the non ergodicity of the 1-ball dynamics [9,10], which is just mentioned in last paragraph, was not described previously too; it has not been found with the help of computers, but its trend is certain, confirmed by a series of experiments and now also by simulations (these ones have to include rotations and solid friction to get the correct trend). So the present result would not be the only case for which computer investigation would be in delay.

*Acknowledgements:* CNES and ESA are thanked for they strong support. for funding the series of parabolic flights in board of the Airbus A300-0g. Assistance from Novespace team has been quite appreciated. The experimental results have been obtained by the team composed of D. Beysens, E. Falcon, S. Fauve, Y. Garrabos, C. Lecoutre, F. Palencia and myself. They are all thanked.

# Feed-back from Readers :

## *Discussion, Comments and Answers*

### From *poudres & grains* articles:

**On *Poudres & grains* 15 (1), 1-16 (2005): remark on 1g *vs.* 0g behaviour of granular gas:** An important argument is missing in the discussion on the difference between 0-g and 1-g experiments on dissipative granular gas in the above quoted paper. Its correction is developed as Remark #9 of section 1 of the next article, but it seems so important that it needs also to be duplicated:

The proposed model finds the probability density function which varies as $f(v)=$ **(A/v)** $\exp(-v/v_o)$ . It works in 0-g because the lifetime $\tau$ of a state "v" scales as L/v in the present model; this generates the 1/v pre-factor in front of the exponential. Applying the same rules in 1g tells that $\tau$ corresponds to the roundtrip time; hence it scales now as 2v/g, which leads to $f(v)=$**(A'v/g)** $\exp(-v/v_o)$. This changes completely the behaviour: it generates a medium with a typical speed; this annihilates the condensation process on the "v=0" state, which is found in 0-g. Hence it makes the physics quite different. To exemplify the difference, let us turn the cell with a single piston upside down, in 1g; this leads to all balls in a condensate at v=0, which demonstrates in turn that the physics at 1g and at -1g are not at all the same. In the same spirit, this forces asking what is the true effect of g-jitter in 0g granular gas? I t may be much more important as thought initially.

<div align="right">P.E.</div>

The electronic arXiv.org version of this paper has been settled during a stay at the Kavli Institute of Theoretical Physics of the University of California at Santa Barbara (KITP-UCSB), in june 2005, supported in part by the National Science Fundation under Grant n° PHY99-07949.

*Poudres & Grains* can be found at :
http://www.mssmat.ecp.fr/rubrique.php3?id_rubrique=402